\documentclass[12pt]{article}
\usepackage{epsfig}
\usepackage{lineno}
\def\beq{\begin{equation}}
\def\eeq#1{\label{#1}\end{equation}}
\def\eeqn{\end{equation}}

\def\beqa{\begin{eqnarray}}
\def\eeqa#1{\label{#1}\end{eqnarray}}
\def\eeqan{\end{eqnarray}}

\let\bar=\overbar

\def\Dslash{\not{\hbox{\kern-4pt $D$}}}
\def\dslash{\not{\hbox{\kern-2pt $\del$}}}

\def\BR{\mbox{\rm BR}}

\def\msb{{\bar{\ssstyle M \kern -1pt S}}}

\def\s#1{\widetilde{#1}}

\input babarsym
\def\babar{\mbox{\slshape B\kern-0.1em{\smaller A}\kern-0.1em
    B\kern-0.1em{\smaller A\kern-0.2em R}}\xspace}

\def\Bp {\ensuremath{B^+}\xspace}

\def\beq{\begin{equation}}
\def\eeq{\end{equation}}
\def\bea{\begin{eqnarray}}
\def\eea{\end{eqnarray}}
\def\bq{\begin{quote}}
\def\eq{\end{quote}}
\def\ben{\begin{enumerate}}
\def\een{\end{enumerate}}

\def \rb {\ensuremath {r_\B}\xspace}

\def \deltab {\ensuremath {\delta_\B}\xspace}

\def\figurebox#1#2#3{%
    \def\arg{#3}%
    \ifx\arg\empty
    {\hfill\vbox{\hsize#2\hrule\hbox to #2{\vrule\hfill\vbox to #1{\hsize#2\vfill}\vrule}\hrule}\hfill}%
    \else
    {\hfill\epsfbox{#3}\hfill}%
    \fi}

\def\Title#1{\begin{center} {\Large {\bf #1} } \end{center}}

\begin{document}
  \linenumbers

\Title{Unitarity Triangle Fitter Results for CKM Angles}

\bigskip\bigskip


\begin{raggedright}  

{\it D. Derkach\index{Derkach, D.}\\
INFN, Sezione di Bologna, I-40127 Bologna, ITALY\\
and\\
LAL, Orsay, F-91898, FRANCE\\
\bigskip 
On behalf of the UTfit collaboration:\\
 A. Bevan, M. Bona, M. Ciuchini, D. Derkach, E. Franco, L. Silvestrini, V. Lubicz, C. Tarantino, G. Martinelli, F. Parodi, C. Schiavi, M. Pierini, V. Sordini, A. Stocchi, V. Vagnoni\\
}

\bigskip\bigskip
\end{raggedright}

Proceedings of CKM 2012, the 7th International Workshop on the CKM Unitarity Triangle, University of Cincinnati, USA, 28 September - 2 October 2012

\begin{abstract}
\noindent
We present the status of the Unitarity Triangle analysis focused on the 
analyses connected to the CKM angles extraction. The angle values are found to be
$\alpha = (90.6\pm 6.6)^{\circ}$, $\sin(2\beta) = 0.68\pm 0.023$, and $\gamma=(72.2\pm 9.2)^{\circ}$.
\end{abstract}

\section{Introduction}

\begin{figure}[htb]
\begin{center}
\epsfig{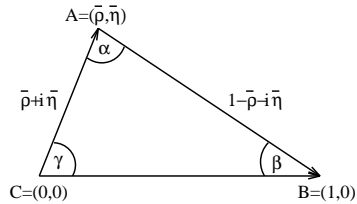}
\end{center}
\caption{Unitarity Triangle in the
 $\bar \rho-\bar \eta$ plane. \label{fig:UT}}
\end{figure}

The Cabibbo-Kobayashi-Maskawa (CKM)
matrix $V_{ij}$~\cite{cite:CKM} has to be unitary, which
implies several relations between its elements. In the 
Wolfenstein parameterizations~\cite{cite:Wolfenstein}, each of these relations can be 
represented as a triangle in the $(\bar\rho, \bar\eta)$ plane. The triangles obtained by product of
neighboring rows or columns are nearly degenerate. The particular interest is driven by the unitarity condition
\begin{equation}\label{eq:uni}
V_{\rm ud}V_{\rm ub}^* + V_{\rm cd}V_{\rm cb}^* + V_{\rm td}V_{\rm
tb}^* =0,
\end{equation}
with each item approximately proportional to $\lambda^3$. This
equation is connected to \B meson decays due to the presence of
$V_{\rm ub}$ and $V_{\rm cb}$ matrix elements. Figure~\ref{fig:UT} 
shows the triangle, which angles, denoted by $\alpha$, $\beta$, and \g,
are\footnote{Another notation for angles, which is also used, is
$\phi_1\equiv\alpha$, $\phi_2\equiv\beta$, and $\phi_3\equiv\g$.
This notation is commonly used by Belle experiment.}:
\begin{eqnarray}\label{eq:angleCKM}
\alpha=\arg\left(\frac{V_{td}^{}V^*_{tb}}{ V_{ud}^{}V^*_{ub}}
\right),
\beta = \arg\left(\frac{V_{cd}^{}V^*_{cb}}{ V_{td}^{}V^*_{tb}}  \right),\\
\gamma = \arg\left(\frac{V_{ud}^{}V^*_{ub}}{ V_{cd}^{}V^*_{cb}}
\right)=\pi-\alpha -\beta.
\end{eqnarray}

These proceedings show the combination of angle measurement and their implementation 
as seen by the UTfit group~\cite{cite:UTFIT}. The combination is performed in the Bayesian approach and
uses the most recent results available by the time of the conference.

\section{CKM angle $\alpha$ extraction}

The CKM angle $\alpha$ is extracted from charmless hadronic \B decays. We use 
the method described in~\cite{cite:UTalpha}. The decays $\B\to\pi\pi$ are
analyzed using the SU(2) isospin symmetry to cleanly disentangle the penguin contribution.
This method relates the isospin amplitudes of $\Bz\to\pip\pim$,  $\Bz\to\piz\piz$, and
$\Bp\to\pip\piz$ processes and
their complex conjugates as two triangles in a complex
plane. We use the \CP-averaged branching fractions of the processes as well as the available 
time-dependent asymmetries. The input values are taken from HFAG~\cite{cite:HFAG}. The same procedure
is applied to the $\B\to\rho\rho$ system with an additional complication of a relative orbital angular momentum. 
A more complicated analysis is used to extract the angle $\alpha$ from the $\Bz\to\rho^0\piz$ decays. Here, we
measure $\alpha$ using a time-dependent Dalitz analysis, which includes the variation of the strong phase of
interfering $\rho$ resonances.

\begin{figure}[htb!]
\begin{center}
\begin{tabular} {c}
\epsfig{file=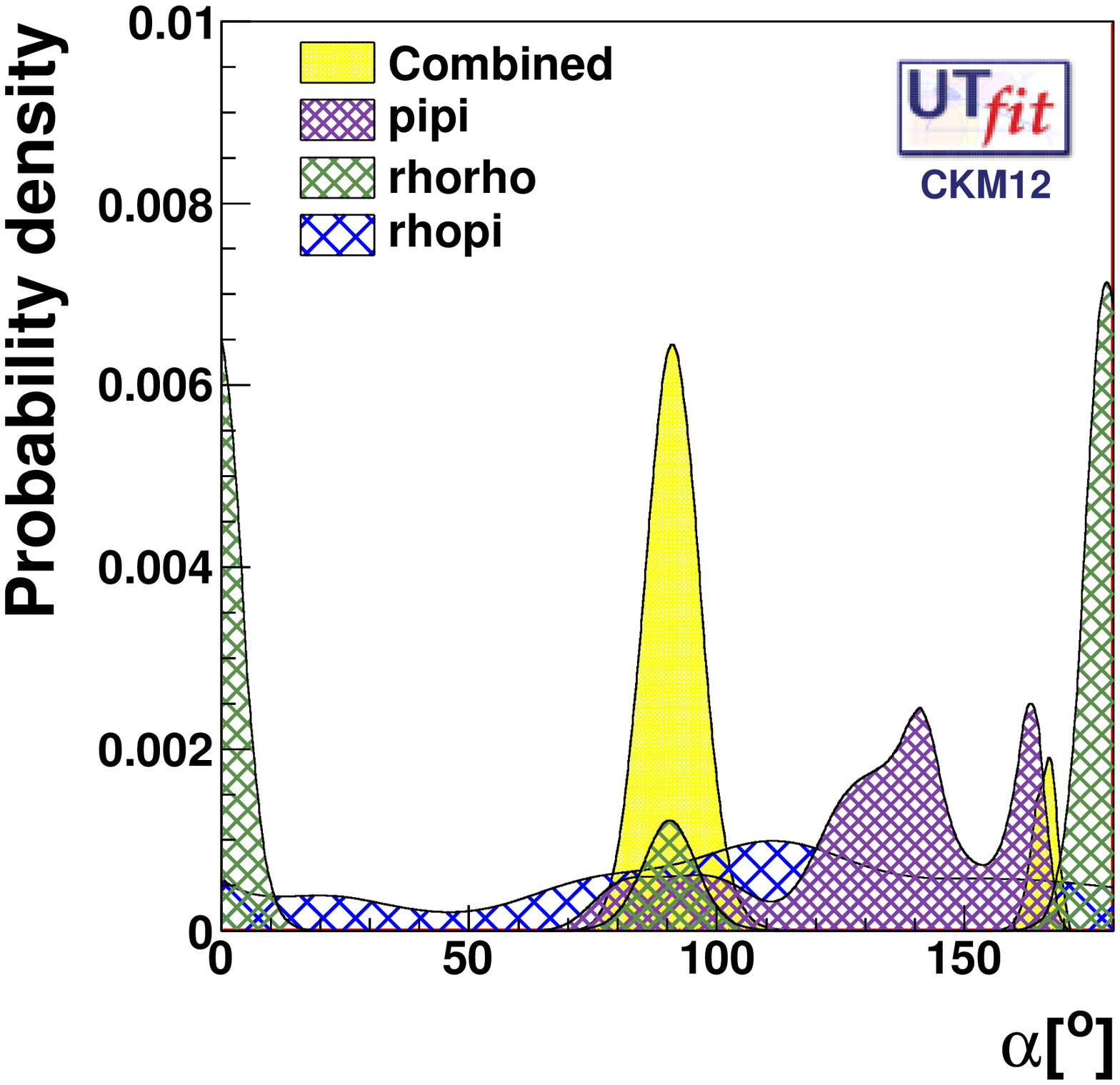,width=0.33\linewidth}
\epsfig{file=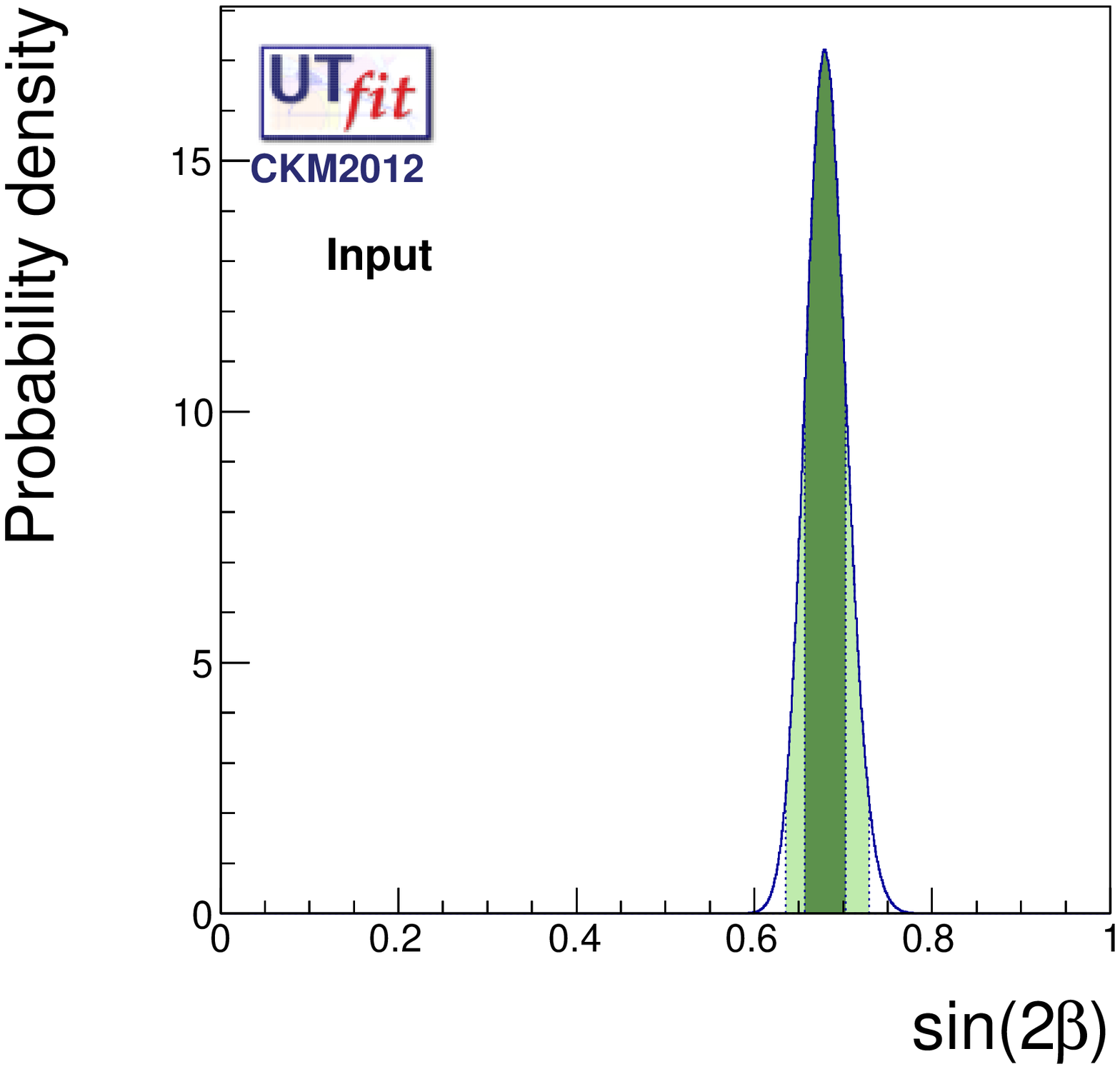,width=0.33\linewidth} 
\epsfig{file=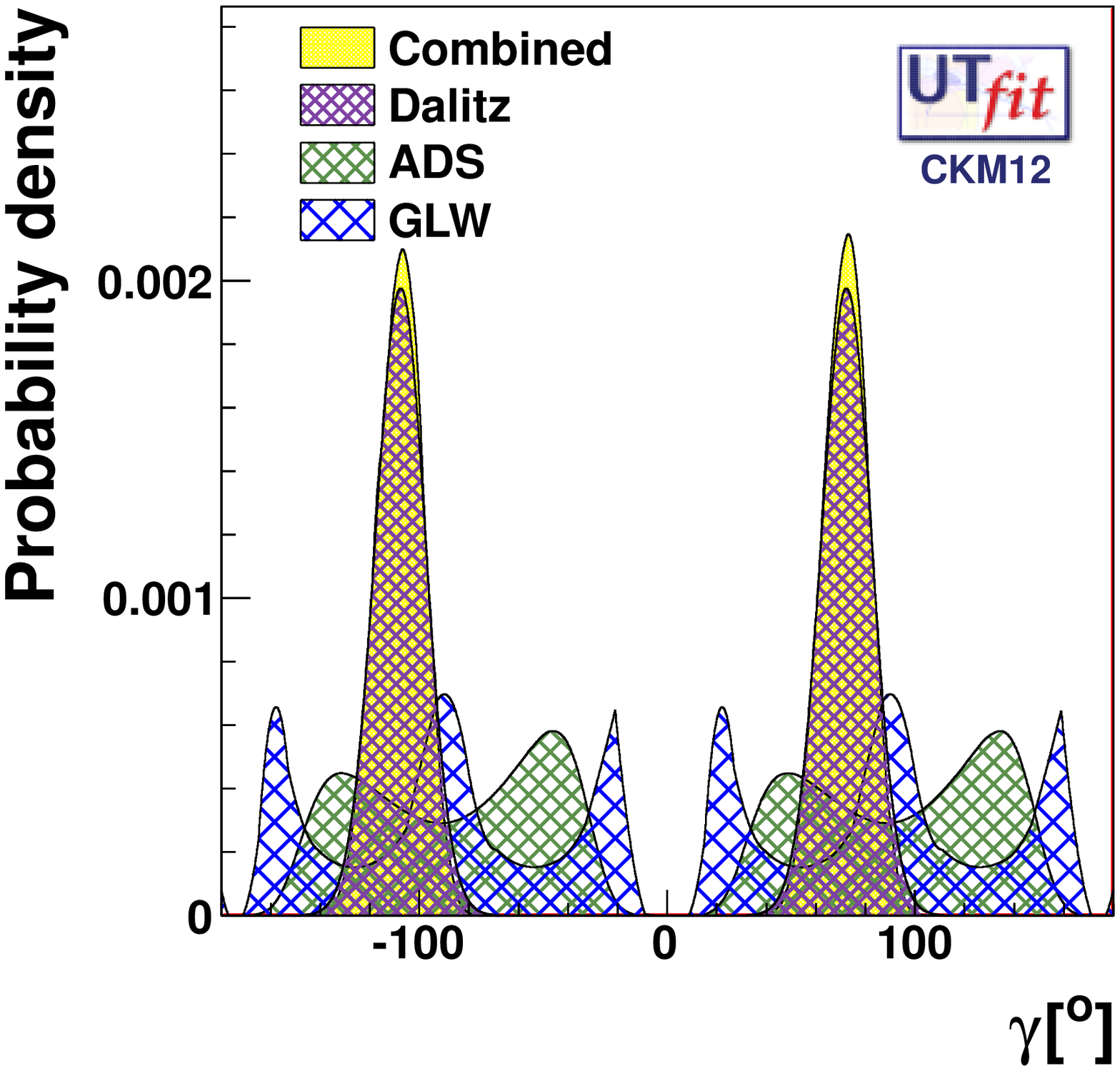,width=0.33\linewidth} 
\end{tabular}
\end{center}
\caption{(color online) One-dimensional probability density functions for $\alpha$ (left), $\sin(2\beta)$ (middle), and $\gamma$ (right) experimental results. 
The plots for $\alpha$ and $\gamma$ also show the contribution from different channels and methods.
\label{fig:angles}
}
\end{figure}

Figure~\ref{fig:angles} shows the combination of the above mentioned methods. This combination gives
$\alpha = (90.6\pm 6.6)^{\circ}$.

\section{CKM angle $\beta$ extraction}

The golden mode to measure the angle $\beta$ is the $\Bz\to J/\psi\Kz$ decay. This mode gives a value of $\sin(2\beta)$
which is considered practically free of theoretical uncertainties and thus serves as a benchmark for
indirect searches for new physics. We estimate the deviation of the measured sine coefficient 
of the time-dependent \CP asymmetry induced by the long-distance contributions from penguin contractions and
by the penguin operators using a data-driven technique~\cite{cite:UTbeta}.  Figure~\ref{fig:angles} shows the combination 
of all the information about the angle $\beta$. This combination gives
$\sin(2\beta) = 0.68\pm 0.023$.
  
\section{CKM angle $\gamma$ extraction}

The CKM angle \g\ is one of the least 
precisely known parameters of the unitarity triangle. The methods of measurements~\cite{cite:dalitz_theo,cite:glw_theo,cite:ads_theo} are 
using charged \B meson decays into $D^{(*)}K^{(*)}$ final states 
which have no penguin contribution. This gives an important difference from most of other direct measurements of the angles. 
These processes are theoretically clean provided that hadronic unknowns are determined from experiment. 
The $\b\to \c\ubar \s$ and
$\b \to \u \cbar \s$ tree amplitudes are used to construct the observables that depend
on their relative weak phase \g, on 
the magnitude ratio $\rb \equiv | {\cal A}(\b\to \u\cbar \s) / {\cal A}(\b\to \c\ubar \s) |$
and on the relative strong phase difference $\deltab$ between the two amplitudes.

The Atwood-Dunietz-Soni method~\cite{cite:ads_theo} needs input from the $D$ meson observables:  amplitudes ratio $r_D$, 
strong phase difference $\delta_D$, and coherence factor $k_D$. 
We perform a fit to the charm sector information 
allowing for \CP violation in the singly-Cabibbo 
suppressed decays~\cite{cite:UTcharm} and receive the following results that are used in the $\gamma$ reconstruction: $\delta_D(K\pi)=(18\pm 12)^{\circ}$
and $\delta_D(K\pi\piz)=(31\pm 20)^{\circ}$. Combining the results obtained by LHCb, BaBar, Belle, and CDF collaborations we obtain $\gamma=(72.2\pm 9.2)^{\circ}$.

The resulting combination is shown in Fig~\ref{fig:angles}. We have also tested the influence of the prior probability distributions and found it to 
be negligible given the statistical uncertainty of the $\gamma$ combination.

\section{Overall Fits}   

\begin{figure}[htb!]
\begin{center}
\begin{tabular} {c}
\epsfig{file=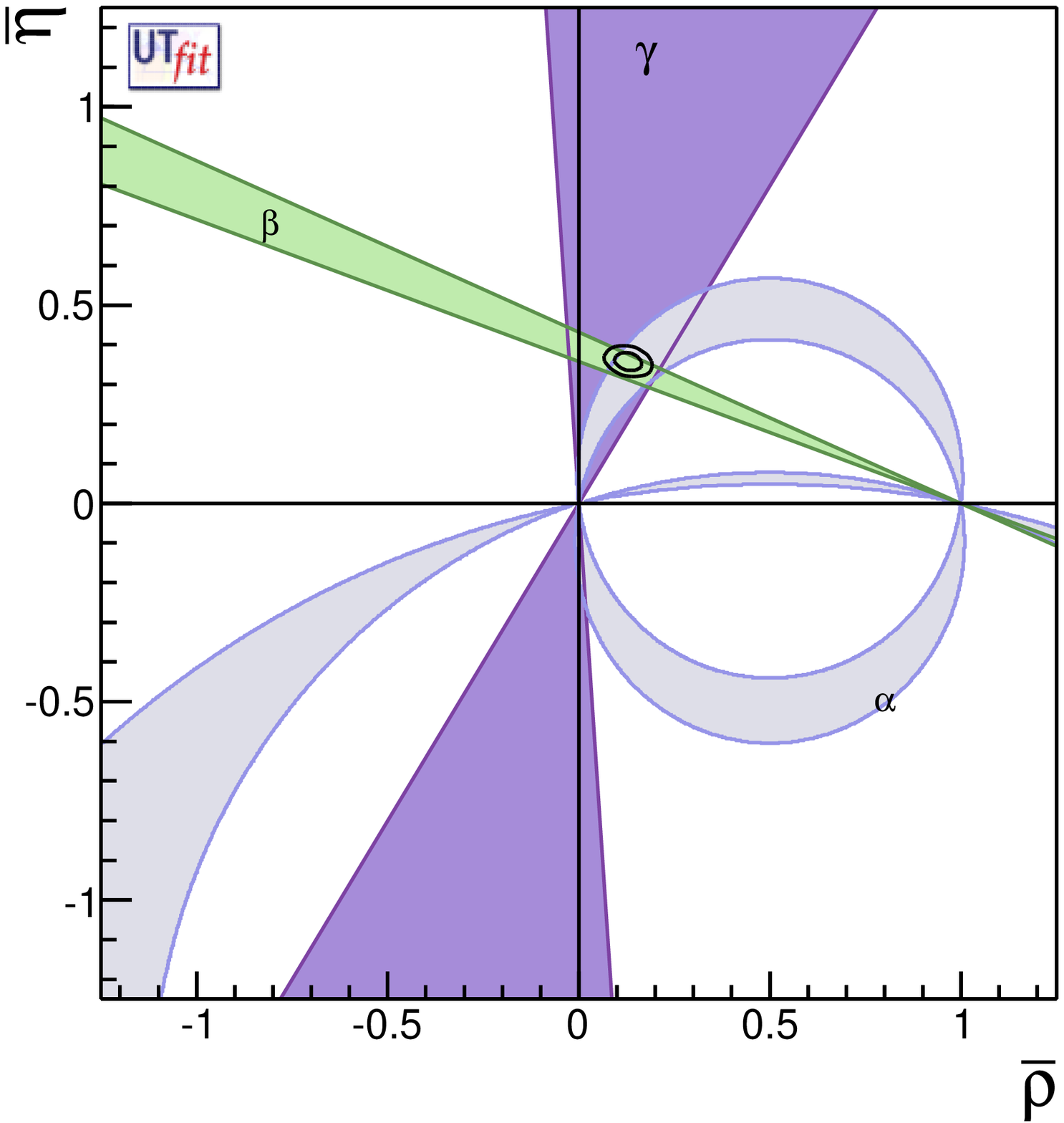,width=0.4\linewidth}
\epsfig{file=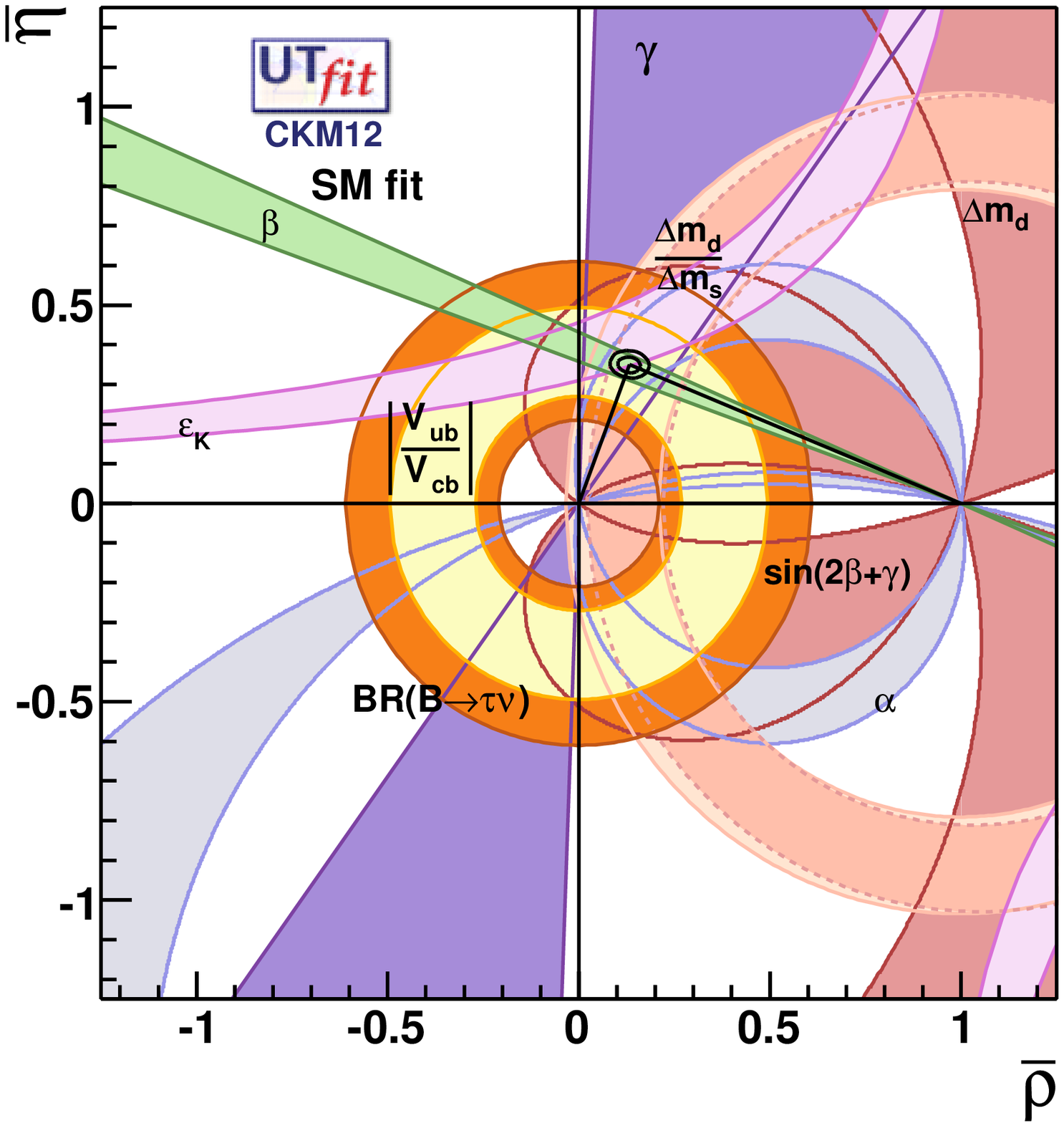,width=0.4\linewidth}
\end{tabular}
\end{center}
\caption{(color online) $\rhobar-\etabar$ planes where the black contours display the 68\% and 95\% probability 
regions selected by the SM global fit. 
The 95\% probability regions selected by the single constraints are also shown.  Left: the “angle-only” fit. 
Right: the global SM fit using all the inputs described in the text.
\label{fig:fullfit}
}
\end{figure}

\begin{figure}[htb!]
\begin{center}
\begin{tabular} {c}
\epsfig{file=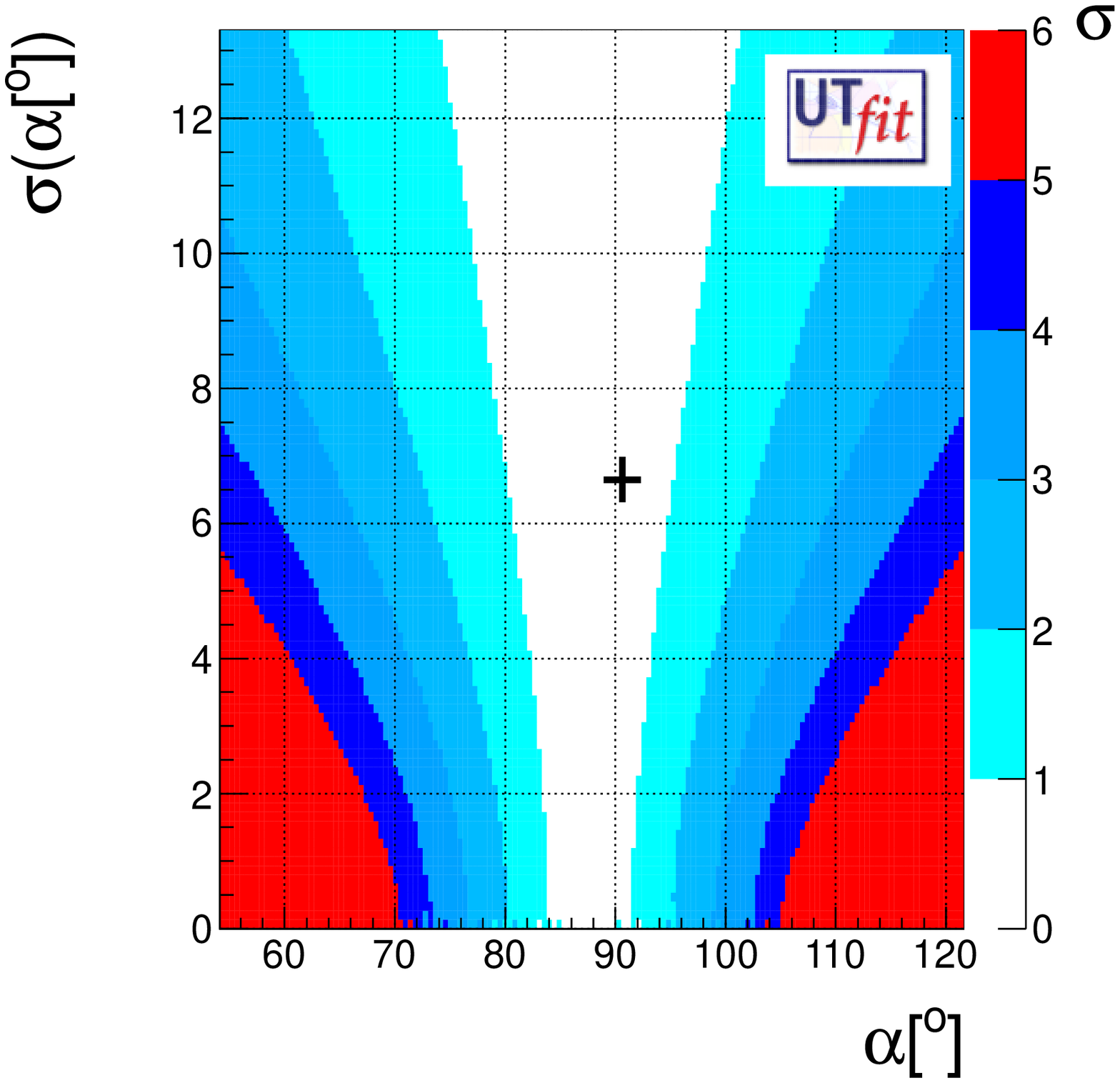,width=0.3\linewidth}
\epsfig{file=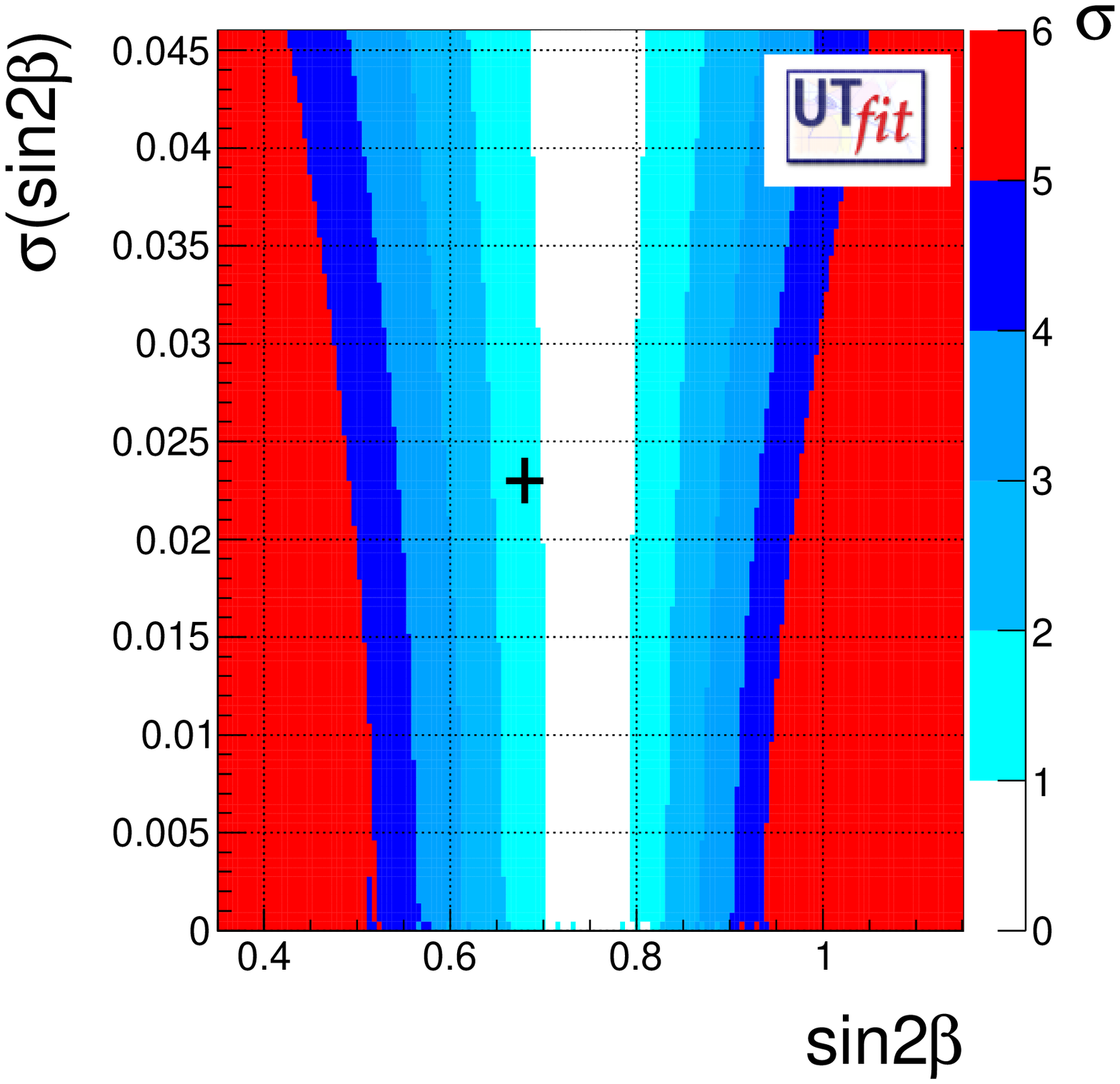,width=0.3\linewidth}
\epsfig{file=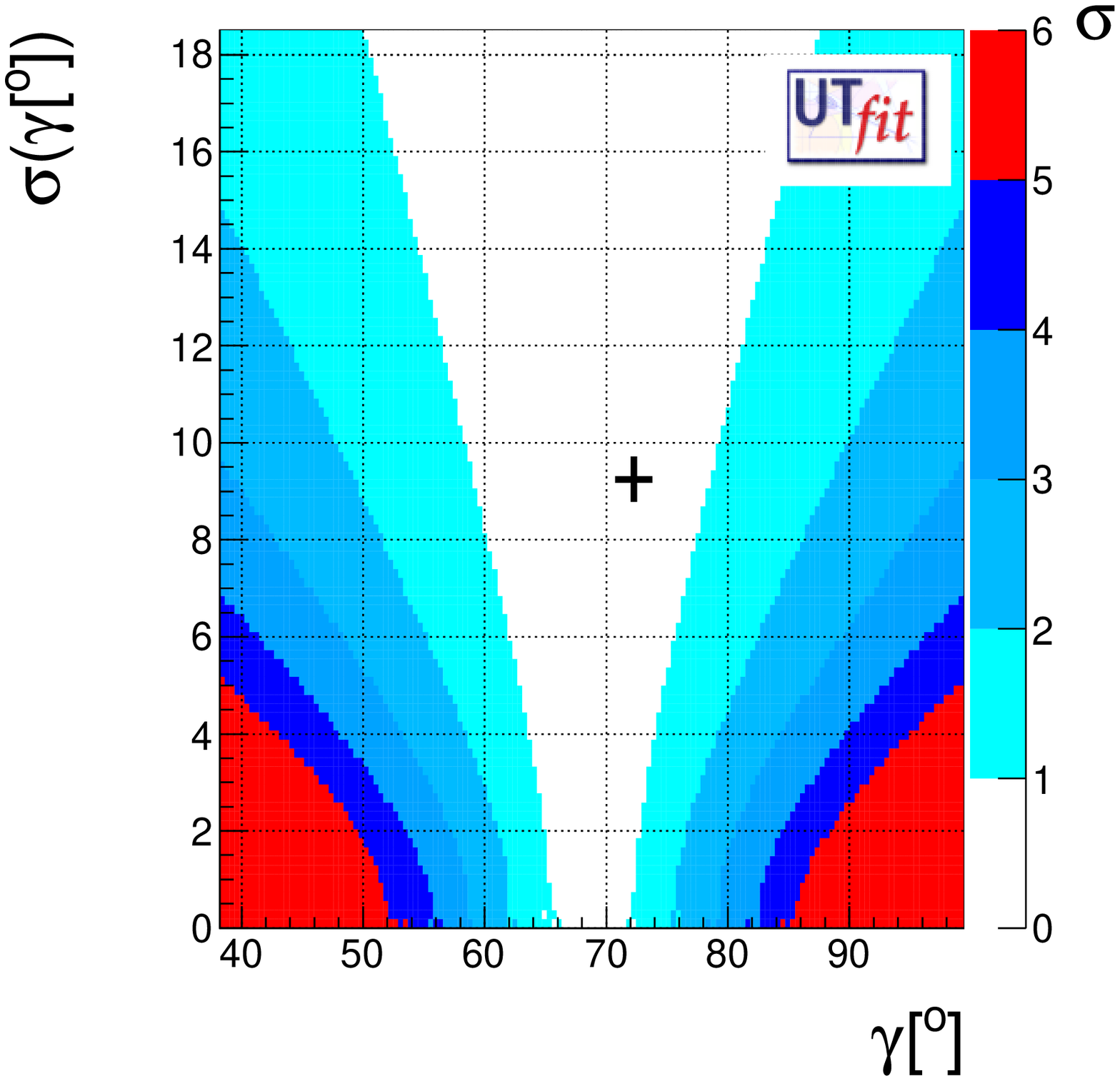,width=0.3\linewidth}
\end{tabular}
\end{center}
\caption{(color online) Compatibility plots for $\alpha$, $\sin (2\beta)$, and $\gamma$. 
\label{fig:pulls}
}
\end{figure}

Using the angle inputs and our Bayesian framework, we perform the fit to the information on angles to extract
the CKM matrix parameters. We obtain $\rhobar = 0.130\pm0.027$ and $\etabar = 0.338 \pm 0.016$.
The resulting fit is shown in Fig~\ref{fig:fullfit}. The fit precision can be improved 
by adding constraints on other parameters: $\Vub/\Vcb$ from semileptonic \B decays, $\Delta m_d$ and $\Delta m_s$
from $B^0_{d,s}$ oscillations, $\epsilon_K$ from $K$ mixing. 
This approach yields $\rhobar = 0.132\pm0.021$ and $\etabar = 0.348 \pm 0.015$. The results of the full fit are shown in 
Fig.~\ref{fig:fullfit}. This approach also allows one to obtain the SM predictions for different observables. The comparisons to the predictions of the 
angle values are shown in Fig~\ref{fig:pulls}. The predictions for the angles are: 
$\alpha = (87.8\pm 3.7)^{\circ}$, $\beta = (24.3\pm 1.9)^{\circ}$, and $\gamma = (68.8\pm 3.4)^{\circ}$.
We do not see big discrepancies between the SM predictions and 
experimental measurements (for more information, see the web-site www.utfit.org).  

\end{document}